%% file: MobileAppUsage_PervasiveHealth.tex
\documentclass[conference]{IEEEtran}
\usepackage{cite}

\usepackage{comment}

\usepackage{todonotes}

\ifCLASSINFOpdf
\usepackage{graphicx}
\else
\fi
\usepackage[cmex10]{amsmath}
\usepackage{url}

\begin{document}
%
\title{Smartphone apps usage patterns as a predictor of perceived stress levels at workplace}


\author{\IEEEauthorblockN{Raihana Ferdous,
Venet Osmani and Oscar Mayora}
\IEEEauthorblockA{CREATE-NET, Trento, Italy\\
\{raihana.ferdous, venet.osmani, oscar.mayora\}@create-net.org}}


%


\maketitle

\begin{abstract}
Explosion of number of smartphone apps and their diversity has created a fertile ground to study behaviour of smartphone users. Patterns of app usage, specifically types of apps and their duration are influenced by the state of the user and this information can be correlated with the self-reported state of the users. The work in this paper is along the line of understanding patterns of app usage and investigating relationship of these patterns with the perceived stress level within the workplace context. Our results show that using a subject-centric behaviour model we can predict stress levels based on smartphone app usage. The results we have achieved, of average accuracy of 75\% and precision of 85.7\%, can be used as an indicator of overall stress levels in work environments and in turn inform stress-reduction organisational policies, especially when considering interrelation between stress and productivity of workers.
\end{abstract}


%
\IEEEpeerreviewmaketitle
\input{introduction.tex}
\input{relatedwork.tex}

\input{methodology.tex}
\input{result.tex}
\input{conclusion.tex}


%
\Urlmuskip=0mu plus 0.2mu
\bibliographystyle{IEEEtran}
\bibliography{referencepervasivehealth}

\end{document}

%% file: introduction.tex
\section{Introduction}
%
%
\label{s:intro}
A number of studies have established that moderate levels of stress can be motivating while, excessive, chronic, and repeated exposure to stress has been associated with chronic diseases, depression, and immune disorders \cite{rosengren2004association, marmot1997contribution}. Prolonged exposure to stress at work has a significant negative effect on work productivity \cite{wilke1985stress}, eventually leading to burnout\cite{maslach2001job}, which reduces the motivation and effectiveness of workers. 
Considering the intertwined relationship between stress, burnout and productivity; monitoring stress at work and understanding related factors is becoming more and more important.
Even though a number of research studies \cite{kyriacou1987teacher, bacharach1991work} have investigated causes and consequences of chronic stress at work through self-reported data, there is far less work focusing on automatic detection of objective behaviour changes. In this regard, mobile technologies can play a significant role in monitoring behaviour of subjects \cite{grunerbl2014smart, matic2014mobile, Matic2012, matic2012speech, maxhuni2015classification}, helped by familiarity of users with these devices in one hand and their multi-modal sensing capabilities on the other.

In this paper we investigate whether it is possible to predict the stress level of the participants considering their app usage patterns. Although there have been studies investigating the usage pattern of mobile apps to predict the next app usage of subjects \cite{do2011smartphone, xu2013preference, liao2012mining, liao2013feature, shin2012understanding}, there has been no study focused on investigating the relationship between the pattern of mobile apps usage of the users at work and their perceived stress levels. 

The results presented in this paper are based on a study of 28 workers, monitored over 6 weeks through their smartphones during their daily, real-world behaviour.
Each participant was provided with a smartphone that had our data collection app installed and there were no restrictions placed upon the subjects as to particular usage or position of the phone.
The purpose of the app was two fold:
i) recording app usage; which is done through a continuous process of storing the active apps, their starting and ending timestamps, and their duration. 
ii) capturing perceived stress levels; which was done through implementation of a questionnaire on the smartphone, where subjects were prompted three times a day to answer a question in order to record their perceived stress level. Initial analysis have revealed that the most active apps, for majority of participants, are email, calendar and browsers; while, the least used apps are games apps, as one would expect in work environments. In total there were 128 unique apps recorded with an average of around 12 apps per participant and the standard deviation is 6.45. While some common characteristics for majority of subjects exist however, the app usage behaviour varies across subjects and it is this variance that we have captured and investigated its relationship with perceived stress levels. In addition, individual differences between subjects have prompted the necessity of devising individual behaviour model for each subject, rather than establishment of a group behaviour model. A classifier is trained on individual subject data and prediction performance is measured for each subject. The classification results are presented in Section \ref{ss:userspecificclassification} and they show that on average, the subjects' stress level can be predicted with 75\% accuracy. The results of the study could be used as an indicator of the overall level of stress in work environments and in turn inform stress-reduction policies, especially when considering interrelation between stress and productivity.

The rest of this paper is structured as follows:
Section \ref{s:relatedwork} covers a discussion on related work of this field, while description of the dataset, feature selection and model development is outlined in Section \ref{s:methodology}.
The experimental results are shown in Section \ref{s:results} and Section \ref{s:conclusion} summarises the results.

%% file: relatedwork.tex
\section{Related Work}
\label{s:relatedwork}
A number of authors \cite{bauer2012can, bogomolov2014pervasive, muaremi2013towards} have investigated correlation between human behaviour and perceived stress at work by analysing the data derived from the smartphones data or sensors data. One of the conclusions is that there are behaviour differences before and during experience of stress, however these differences are distinct for each individual. Several studies \cite{bohmer2011falling1, falaki2010diversity, xu2011identifying} investigating the pattern of smartphone apps usage have noticed immense diversity among smartphone users in different aspects, such as, per day user interaction, application use, network traffic and energy drain. 
The authors of \cite{falaki2010diversity} have suggested to learn and adapt user specific interaction pattern with smartphones in order to improve user experience and to predict the energy consumption more accurately. A number of research papers \cite{ liao2012mining, liao2013feature, shin2012understanding, huang2012predicting, bohmer2013appfunnel, Gruenerbl2014} are found performing large scale analysis of the context related to mobile app use. Here, the authors have noticed strong dependencies between app usage and the contextual variables, such as, location, time and social aspects. They have shown that the rich contextual information related to the smartphone apps usage has the power to predict the next app that will be used by the user. Authors of \cite{chittaranjan2013mining} have shown that aggregated features obtained from smartphone usage data (e.g., app usage, phone call records, location information, etc) are indicators of the Big-Five personality traits (i.e., extraversion, agreeableness, conscientiousness, emotional stability and openness to experience). 
Authors of \cite{alvarez2014tell} have reported that the changes in general behaviour of patients due to onset of a bipolar episode can be captured through the differences in their smartphone usage pattern, while a study investigating the stress and strain of student life is reported in \cite{wang2014studentlife}.

%% file: methodology.tex
\section{Methodology}
\label{s:methodology}
\subsection{Collection of Data}
\label{ss:datacollection}
Each of the 28 participants was provided with a smartphone, having our data collection application installed and were monitored over a period of 6 weeks, from November 2013 to December 2013.
As our focus is to investigate the participants' behaviour during work, we limit the collection of app usage records only for the duration they spend at work. 
The timestamped records of start and end times of each active applications used by the participants are captured by our data collection app. 
In order to identify which applications are used by the participant, we considered those applications that are running within the time period when the screen status is on. 
In order to understand the subjective stress levels of the participants, a question (``what is your stress level?'') was answered on a 5-point scale, where 1 indicates ``very slightly or not at all'' and 5 indicates ``extremely''.
This questionnaire was implemented on the smartphones of the participants and was asked 3 times during the day, in the morning at the beginning of the work, around noon, and before leaving workplace.
\subsection{Feature Extraction \& Prediction of Stress}
\subsubsection{Categorisation of Apps}
In order to get a more high level understanding of the applications used by the participants during the observation window, we have divided apps into 5 categories: (i) entertainment, (ii) social networking, (iii) utility, (iv) browser, and (v) game apps. 
Table \ref{tab:categorizationofapps} presents a number of examples of apps for each category.
\begin{table}
\centering
\caption{Categorisation of apps used by the participants}
\label{tab:categorizationofapps}
\begin{tabular}{ |p{3.8cm}|p{4cm}| }
  \hline
 {\bf Social networking service apps} & {\bf Entertainment applications} \\ \hline  
		Facebook, Twitter, google+,    & Audio \& Video streaming apps,\\ 
      VoIP apps & Online book reading and purchase,\\     
    (e.g., LocalPhone, MobileVoIP, etc.), & FM Radio, tv guide, \\
     Dropbox, Pinterest, Flipboard,& Music instrument tuner \& composition, \\    	
		Whatsapp, gtalk, viber, skype& Youtube, video player, tv streaming,\\     	
		& News listening and reading apps\\ \hline   	
 {\bf Utility applications} & {\bf Browser applications} \\ \hline
		Calendar, map, navigator, clock,   & Chrome \\ 
    Weather app, note taking app,  & Firefox \\     
    Voice to text, text to voice app,   & Email \\
    Trip advisor, QR Code Reader,  &   google quick searchbox\\     		
		Calculator, accounting app, &   \\
		Scanner, flashlight, voice recorder,&   \\
		Recipe apps, camera, Online shopping  &   \\ \hline    		
 \multicolumn{2}{|c|}{\bf Game applications} \\ \hline
 \multicolumn{2}{|l|}{Card games, puzzles, fun games (e.g., fruitninja, candycrushsaga, angry birds)} \\ \hline
\end{tabular}
\end{table}
\subsubsection{Feature Selection}
In order to characterize the subject-centric app usage patterns, we have extracted a set of suitable features from the collected app usage records of the participants. 
Smartphone app usage behaviour is driven by factors such as popularity of apps (e.g.,  how many categories of apps a user visited and how is his/her attention spread), frequency and duration of app usage (e.g., how often does a user uses any application and how long is the app used), frequency of unique apps used (e.g., if the user keeps using only a few apps regularly or the list of his/her visited apps is large). 
Considering above criteria, a set of features (shown in Table \ref{tab:featurelist}) is extracted from the app usage records of each participants that captures their usage pattern per day.
\begin{table}
\centering
\caption{List of features extracted from the app usage records of a user}
\label{tab:featurelist}
\begin{tabular}{|p{2cm}|p{6cm}|} 
    \hline  
        {\bf Feature} & {\bf Description}\\ \hline
    Frequency ent & Frequency of using apps belong to {\it entertainment} category per day\\ \hline
		Time ent & Time (in second) of using apps belong to {\it entertainment} category per day\\ \hline
		Frequency social & Frequency of using apps belong to {\it social networking service} apps category per day\\ \hline
		Time social & Time (in second) of using apps belong to {\it social networking service} apps category per day\\ \hline
		Frequency game & Frequency of using {\it game} apps per day\\ \hline
		Time game & Time (in second) of using {\it game} apps per day\\ \hline
		Frequency utility & Frequency of using apps belong to {\it utility} category per day\\ \hline
		Time utility & Time (in second) of using apps belong to {\it utility} category per day\\ \hline
		Frequency browser & Frequency of using apps belong to {\it browser} category per day\\ \hline
		Time browser & Time (in second) of using apps belong to {\it browser} category per day\\ \hline
		Number of unique app  & Number of unique apps used per day \\ \hline
		\end{tabular}
\end{table} 

\subsubsection{Prediction of stress level of a user}
User-centric model of app usage is developed by training a SVM classifier with individual app usage behaviour (features extracted from the app usage record) and participants' perceived stress levels as ground truth.
The model thus trained with the user-specific app usage information is successfully able to predict the stress level of that user during the testing period.
%
%
Before training each classifier, a suitable model (i.e., kernel function and the SVM parameters) is selected. 
%
%
Selection of kernel and tuning of the parameters of SVM are performed on the training set through the widely used  {\em cross-validation technique}. 
This test was performed for three kernel functions: linear, RBF, and polynomial -- and for various values of their relevant parameters (for example, SVM parameter {\em soft margin constant} $C$ to controls the trade-off between maximizing the margin and minimizing the training error, RBF kernel parameter $\gamma$, and, the degree $d$ for polynomial kernel). 
%
%
Cross-validation method is applied for each set of parameters. 
Finally, the parameters with the best cross-validation index were selected. 

%% file: result.tex
	\section{Analysis and Experimental Results}
	\label{s:results}
	\subsection{General pattern of smartphone app usage}
\begin{figure}[h!]
	\centering
	\includegraphics[scale=0.6]{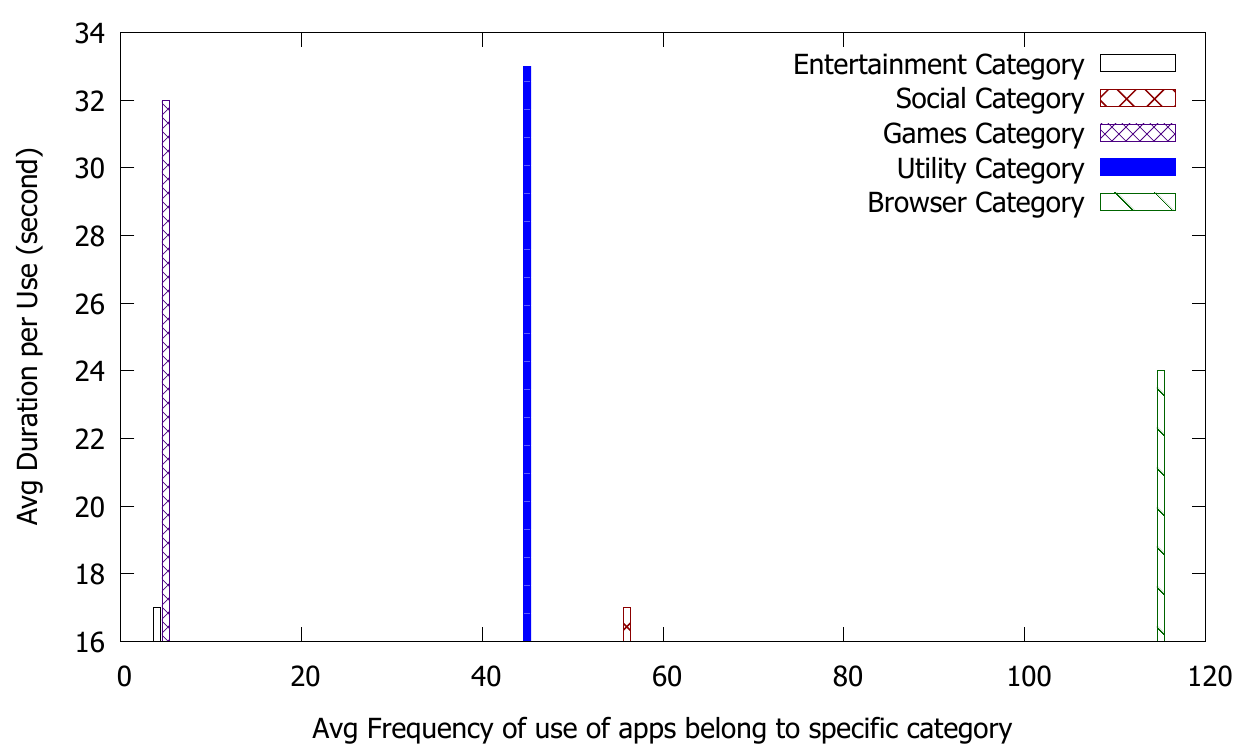} 
	\caption{Average number of times apps of specific categories are used vs duration per use}
	\label{fig:appfrequencyandduration}
	\end{figure}

\begin{figure}[h!]
	\centering
	\includegraphics[scale=0.72]{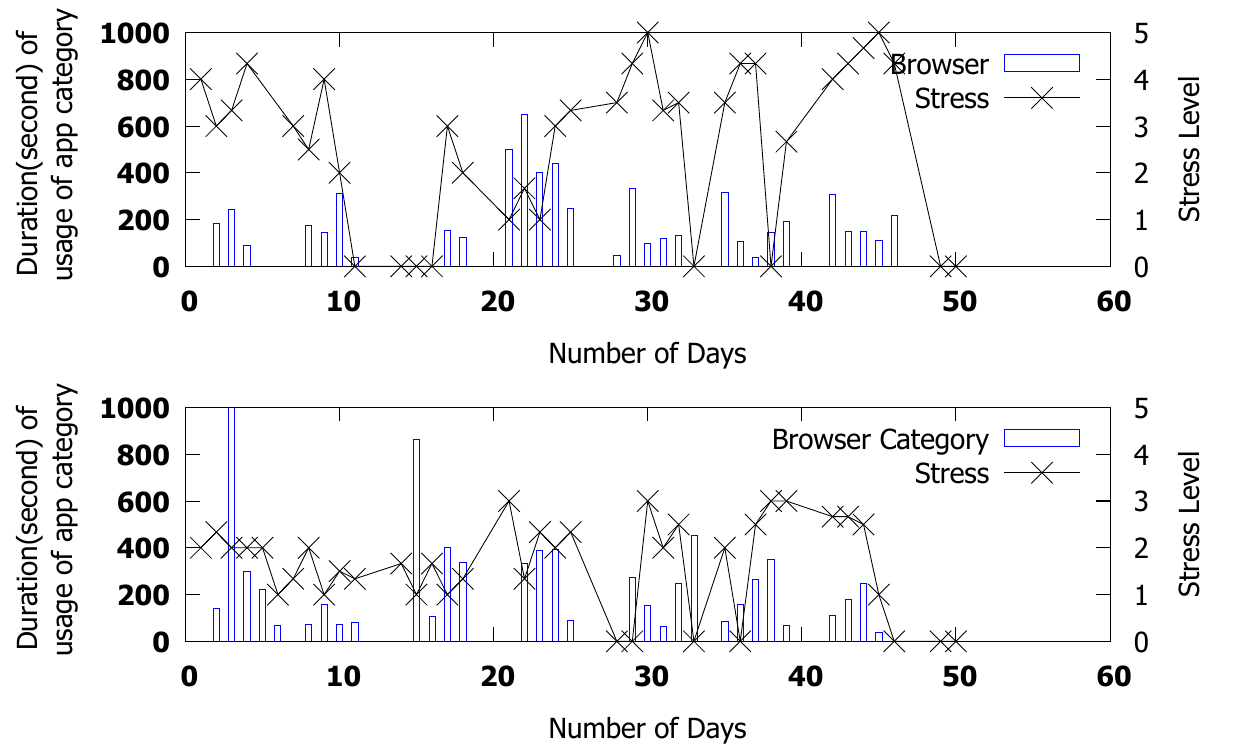} 
		\caption{Usage pattern of apps belonging to {\it Browser} category and stress level (1-5) per day for two randomly chosen participants [stress level, 0 = no data available]}
	\label{fig:user2appusage}
	\end{figure}

A number of common characteristics are observed across all the participants regarding their app usage history. 
Figure \ref{fig:appfrequencyandduration} presents the relationship between average usage time and frequency for applications of each category.
It can be seen that apps belonging to the {\it browser} category have high usage frequencies but lower durations per use across all participants. Apps belong to {\it utility} category are also used frequently and for a longer duration. 
This is something that is expected in a working environment where use of apps such as email, calendar, browser, is common. 
Interestingly, in case of games apps, the average duration is high, however they are used few times only. This may be due to the nature of games, where once the users begin a game they typically stay engaged for longer duration, in comparison to say calendar, browser or email apps.
	In the context of perceived stress at work, the relationship between the app usage patterns and perceived stress also varies for different participants. For instance, Figure \ref{fig:user2appusage} shows the diversity of using apps belonging to {\it browser} category and stress level of two randomly chosen participants. 
    
	\subsection{Generic model for stress level prediction}
	\label{ss:genericclasifier}
	Considering the common characteristics of app usage observed for majority of the participants, first we have focused on developing a generic group behaviour model to predict the stress level.  
	To this end, a combined dataset is prepared by integrating the features extracted from the app usage records for all the participants. 
	The model is developed by training a SVM classifier with a subset of this dataset defined as {\it training} data, while the remaining subset is used for testing. 
	Here, k-fold (k=10) is used, which indicates that each subset is tested (validated) using the model trained on the remaining $\left(k-1\right)$ subsets with the right choice of kernel function and SVM parameters. 
	The $k$ results are averaged to obtain a single index. 
	In this context, the average accuracy obtained from the generic classifier is 54\%. This result indicates the limitations of using group behaviour model in predicting stress level of the participants considering their app usage history.
    \subsection{User-centric model for stress level prediction}
    \label{ss:userspecificclassification}
    Considering the diversity of individual in the context of app usage, we have focused on developing user-centric model to find the correlation between app usage and perceived stress level. 
	Here, each model is trained with the individual app usage features and his/her perceived stress level.  
To this end, the app usage features of each user is divided into two sets: (i) training set consists of 70\% data, and (ii) test set containing the remaining 30\% data. 
User specific models are selected by using k-fold cross-validation technique on the training sets. 
The validation of models is performed on the test sets. 
Table \ref{tab:result} shows the stress prediction accuracy obtained through the 10-fold cross-validation during model selection process from 22 models that correspond to 22 active participants (out of 28 participants that took part in the study). While the highest accuracy is 99\% and lowest accuracy is 69.2\%, the table shows that an average accuracy is 88.1\%. 
The third column of the Table \ref{tab:result} shows the stress prediction accuracy obtained from the user specific models during the validation process. While the highest accuracy is 95\% and lowest accuracy is 50\%, the table shows that an average accuracy is 75\%. 
	

    
   
	%
	%
\begin{table}
\centering
\caption{Accuracy, precision, and recall for all the participants}
\label{tab:result}
 \begin{tabular}{ |c|c|c|c|c|}
    \hline
    User ID& \vtop{\hbox{\strut Model Selection Accuracy}\hbox{\strut (Cross-Validation)}} & \vtop{\hbox{\strut Accuracy}\hbox{\strut (Validation)}} & Precision  & Recall \\ \hline 
    84616 & 74.194\% & 60\%& 72.727\% & 88.889\%  \\ \hline
	94433 & 81.481\% & 66.67\% & 85.714\% & 60.000\% \\ \hline
	88187 & 77.273\% & 57.14\% & 72.727\% & 80.000\%  
\\ \hline
	89532 & 86.957\% & 57\% & 88.889\%	 & 80.000\% \\ \hline
	94441 & 96.552\% & 80\% & 100.000\% & 92.308\%  
 \\ \hline	
	95646 &91.304\% & 80\% & 84.615\% & 85.714\%  
\\ \hline		
	14446 & 83.871\% & 60\% & 84.615\%	 & 78.571\%  
\\ \hline		
	96479 & 69.231\% & 50\% & 58.333\% & 	70.000\%  
\\ \hline		
	94516 & 75.000\% & 58\%& 69.565\% & 	100.000\%  
\\ \hline		
	94813 & 99.000\% & 90\%& 98.000\%	 & 96.000\%  
\\ \hline		
	94615 & 93.750\% & 80\%& 50.000\%	 & 100.000\%  
\\ \hline		
	95414 &94.444\% & 91\%& 100.000\%	 & 50.000\%   
\\ \hline		
	95596 & 95.000\% & 88\%& 91.435\% & 	93.000\%  
\\ \hline		
	95448 & 75.758\%& 70\% & 53.333\%	 & 88.889\%  
\\ \hline		
	87676 & 88.000\%& 87.50\% & 88.374\%	 & 74.000\%  
\\ \hline		
	93401 & 96.154\% & 75\%& 100.000\% & 83.333\%  
\\ \hline		
	95216 & 94.737\%  & 83.33\%& 100.000\% & 50.000\%
\\ \hline		
	87684 &98.000\% &	95\% & 96.000\%	 & 92.000\%  
\\ \hline		
	94714 & 86.667\% & 55\% & 100.000\% & 60.000\% 
\\ \hline		
	89953 & 85.714\% & 71.42\%& 100.000\% & 63.333\%  
 \\ \hline		
	96040 & 98.000\% & 95\% & 95.000\%	 & 92.545\%  
 \\ \hline		
	95521 & 98.000\%  & 93\%& 96.000\% & 	97.594\% 
\\ \hline	
    {\bf Average} & {\bf 88.140\% }&  {\bf 75\%
}& {\bf 85.697\% } &  {\bf 80.735\%}\\ \hline
   \end{tabular}
\end{table} 

%% file: conclusion.tex
\section{Conclusion}
\label{s:conclusion}
The results presented in this paper have shown that patterns of smartphone app usage are highly correlated with the self-reported stress level of the users and consequently can be used to predict stress levels within the workplace context. While there is some variation across participants, on average the results we have achieved can be used as solid indicators of individual and organisation-wide stress level, without requiring any intervention from the users. Furthermore, monitoring and understanding stress level has wide-reaching implications in informing organisation of workplaces.

